\documentclass[fleqn,10pt]{wlscirep}
\usepackage{graphicx}
\usepackage[T1]{fontenc}
\usepackage{bm}
\graphicspath{{Images/}}  
\usepackage{multicol}
\usepackage{float}
\usepackage{comment}
\usepackage{amsmath}
\usepackage[misc]{ifsym}
\usepackage{textcomp}

\begin{document}
\title{DNA Ternary Full Adder}

\author[1]{Enqiang Zhu}
\author[1]{Peize Qiu}
\author[2 $^{\textrm{\Letter}}$]{Xianhang Luo}
\author[3 $^{\textrm{\Letter}}$]{Chanjuan Liu}
\author[4]{Jin Xu} 
\affil[1]{Institute of Computing Science and Technology, Guangzhou University, Guangzhou 510006, China}
\affil[2]{School of Computer Science and Technology, Wuhan University of Science and Technology, Wuhan 430065, China}
\affil[3]{School of Computer Science and Technology, Dalian University of Technology, Dalian 116024, China}
\affil[4]{School of Computing Science, Peking University, Beijing 100871, China}

\affil[$^{\textrm{\Letter}}$]{Corresponding authors: 202313601008@wust.edu.cn; chanjuanliu@dlut.edu.cn}

\begin{abstract}

As transistor dimensions continue to shrink, binary devices are rapidly approaching their fundamental limits in power density. In response, multi-valued systems have attracted significant attention due to their enhanced information density. Among these, the ternary system stands out as the most practical option, being the closest integer base to \(e\), which is considered optimal for information efficiency. Despite the intrinsic advantages of DNA nanomaterials, such as programmability, energy efficiency, and massive parallelism, their application in ternary logic remains largely unexplored, particularly in the realm of ternary addition circuits. This gap can be attributed to a fundamental challenge: ternary logic requires circuits capable of recognizing and processing a far larger set of input combinations than binary systems, a task that existing models and techniques often struggle to accomplish. In this work, we propose a novel architecture for a ternary full adder. Our design includes a competitive blocking (CB) circuit that enables the recognition and computation of all possible three-input ternary combinations. Coupled with a dynamic concentration adjustment (CA) strategy, this approach significantly enhances the number of trits that can be processed. Biochemical experiments demonstrate that the CB circuit successfully yields the correct output digits for a ternary full adder, achieving 17-trit ternary addition. To our knowledge, this work represents the first successful DNA-based ternary adder, establishing a new methodological foundation for DNA computing and highlighting its considerable potential for scalable digital information processing.

\vspace{0.1cm}
{\bf Keywords:}  DNA ternary adder; competitive blocking circuit; concentration adjustment.
\end{abstract}

\flushbottom
\maketitle

\section*{Introduction}


Molecular systems represent a revolutionary intersection of biology, computing, and engineering, offering unique capabilities that traditional silicon-based systems cannot match \cite{goni2019high}. By leveraging the Watson-Crick base-pairing principle, DNA molecules exhibit remarkable programmability at the molecular level, enabling them to evolve into a novel computational paradigm. Since Adleman \cite{adleman1994molecular} first employed DNA to solve the traveling salesman problem, DNA computing has attracted significant attention. DNA molecules are no longer seen merely as a carrier of genetic information; it is increasingly utilized as a sophisticated engineering material across various research fields \cite{fan2020propelling}, including the development of DNA nanostructures \cite{kim2018detection,gasser2017sensing,spencer2021binding}, the construction of DNA logic gates \cite{jingjing2023three,zhou2016multifunctional,zhang2023implementing}, and applications in disease detection \cite{allemani2018global,zandvakili2019cell,shieh2020advances,yang2021dna}.  DNA computing offers remarkable advantages in parallel processing, which greatly enhances computational efficiency \cite{chen2023dna,xu2022graph}.

\setlength{\parindent}{1em}The toehold-mediated strand displacement (TMSD) plays a crucial role in DNA computing due to its unique properties. Firstly, it operates spontaneously and stably at room temperature without requiring enzymes \cite{liu2017regulating,oishi2018comparative}. Secondly, TMSD demonstrates precise sequence orthogonality, enabling specific interactions among different sequences. This characteristic is essential for constructing complex computational networks, thereby significantly enhancing computational accuracy and reliability \cite{lv2021biocomputing,wang2023parallel}. By leveraging TMSD technology, researchers have successfully created a range of molecular computing devices \cite{seelig2006enzyme, elbaz2010dna, orbach2014full, liu2020cross, lv2023dna, wang2020implementing, weng2022cooperative, wu2024plug}. Among them, the “seesaw” gates introduced by Qian and Winfree were the first to realize large-scale digital circuits\cite{qian2011scaling} and neural networks\cite{qian2011neural}, and are now regarded as the cornerstone of scalable strand-displacement logic. Meanwhile, DNA tile-based computing \cite{evans2017physical, woods2019diverse, sarraf2023modular} and TMSD circuits on top of DNA origami \cite{amir2014universal, chatterjee2017spatially, bui2018localized, li2025programming} have further spatialized and structured computational units, expanding the design dimensions. In addition, some studies have focused on optimizing the TMSD reaction process, including suppressing leakage \cite{song2018improving}, enhancing signal output \cite{turberfield2003dna, zhang2007engineering}, and tuning reaction kinetics \cite{zhang2009control}.  These advances provide important theoretical support for the field of DNA computing. 

Among the various molecular computing devices, the prominence of addition operations is particularly pronounced. Addition serves as the cornerstone of a myriad of digital arithmetic operations, including magnitude comparison \cite{mehra20132}, multiplication \cite{imana2017fast,navi2009two}, and subtraction \cite {kouretas2012low}. Furthermore, it constitutes a core component of arithmetic modules within digital microprocessors \cite{aguirre2010cmos}, where its performance directly dictates the overall efficiency of the computing system. Consequently, DNA adders, which exemplify the construction of fundamental arithmetic operations from DNA logic gates, demonstrate TMSD's profound potential for implementing intricate logical functions at the molecular level, thereby establishing a foundation for DNA computing models for more sophisticated tasks \cite{tang2025localized}.

\setlength{\parindent}{1em}Considerable efforts have been devoted to building DNA adders. Early logic gates primarily utilized binary encoding, employing single-stranded DNA as input and optical or electrical signals as output \cite{li2015implementation,han20188}. These pioneering efforts successfully constructed fundamental logic gates, including XOR and AND gates \cite{park2012simple}. Then, Li et al. \cite{li2016dna} combined these XOR and AND gates to develop a half-adder. However, its complex logic-gate design, which exhibits low integration efficiency, makes the DNA full adder difficult to implement. For this, Su et al. \cite{su2019high} applied dual-rail DNA logic gates by paralleling single-rail AND and OR gates, effectively integrating a 1-bit full adder with a 4:1 multiplexer to realize a vital DNA arithmetic logic unit. To further simplify circuit complexity, Wang et al. \cite{wang2020implementing} proposed a strategy based on the principle that ``any digital circuit can be represented by a set of equations, with equation terms corresponding to specific switches in the circuit.'' This approach employed TMSD to create DNA switch circuits for molecular digital computing, using only 20\% of the DNA strands compared with dual-rail logic expressions. In addition, some studies have achieved DNA adders by introducing complex structures, such as DNA tetraplexes \cite{huang2019versatile} and DNA origami \cite{tang2025localized}. However, many molecular designs within these structures face scalability issues that impede effective signal transmission. Although Tandon et al. \cite{tandon2020demonstration} designed a DNA tile-based calculator capable of performing addition on 6-bit binary numbers, its experimental operation and system design are more complex. To overcome these challenges, Xie et al. \cite{xie2022scaling} designed a scalable DNA full adder based on cooperative strand-displacement reactions. By developing a dual logic gate capable of simultaneously performing XOR and AND operations, they significantly reduced the number of DNA strands required for the computation. Furthermore, Stérin et al. \cite{sterin2025thermodynamically} implemented the addition of up to 25 bits by combining thermodynamics and self-assembly techniques.


However, existing DNA adders are all based on binary encoding. From the perspective of information economy, the e-ary system is superior (see Supporting Information, Section~S3). Specifically, when representing numbers over the same range, the product of base and width for the e-ary system is smaller than that for binary, meaning that under ideal conditions the e-ary system can achieve the same computational scale with fewer resources~\cite{hayes2001third}. The e-ary system is closest to the ternary system. Moreover, studies have shown that when binary Boltzmann transistors in traditional CMOS technology approach the power density limit, the most effective way to significantly increase information density ($N$) is to switch from a binary system to a ternary system, where the system complexity is reduced to $C = \log_3 N / \log_2 N \approx 63.1\%$ \cite{jeong2019tunnelling}. Unfortunately, ternary logic faces an inherent challenge: by its nature, it requires the circuit to recognize and correctly compute a far greater number of input combinations than binary systems, a requirement that cannot be met with existing models or techniques. Consequently, a ternary adder has not yet been experimentally realized in DNA computing.
\begin{figure*}[t]
    \centering
    \includegraphics[width=1\linewidth]{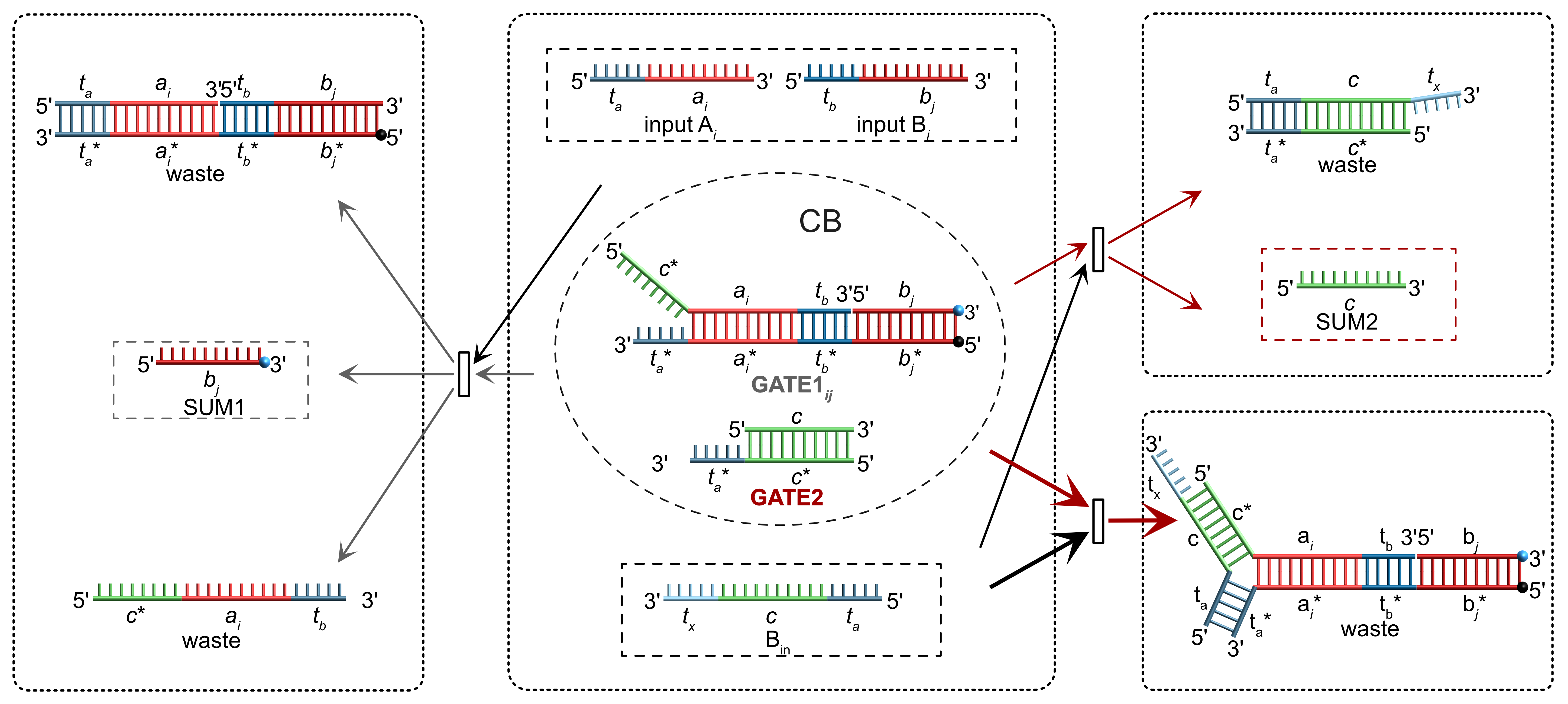}
    \caption{\small The illustration depicts the operational mechanism of the CB circuit, which encompasses three core reactions corresponding to Equations 1, 2, and 3. All reactions involving GATE$1_{ij}$ are represented by gray arrows, while those involving GATE2 are indicated with red arrows. It is particularly noteworthy that the reaction efficiency between GATE$1_{ij}$ and B$_{in}$ is the highest. To emphasize this significant difference in efficiency, we have intentionally employed a thicker red arrow to represent the reaction between GATE2 and B$_{in}$.}
    \label{figure1}
\end{figure*}

\setlength{\parindent}{1em}To fill this gap, this paper proposes a DNA-based ternary adder architecture. Full adders require flexible handling of carry information, a challenge necessitating precise chemical reaction network design. To address this, our study designed a competitive blocking (CB) circuit to accurately recognize and manage the carry information from prior calculations along with the two current addend information (i.e., a total of three inputs, giving 18 possible combinations). The core of this circuit lies in its clever utilization of differences in reaction rate constants (e.g., $k_{2} \gg k_{1}, k_{3}$ in reactions 1, 2, 3) to enable dynamic selection and blocking of reaction pathways, essentially functioning as a logic control based on chemical kinetics. Biochemical experiments demonstrate that the CB circuit can reliably determine whether a carry occurred in prior computations and rapidly respond by selecting the corresponding reaction path. Furthermore, this research, for the first time, successfully integrated ternary logic with the proposed concentration adjustment (CA) strategy and, based on the CB circuit, designed and implemented a DNA ternary full adder. The CA strategy aims to minimize the demand for carry information during computation, thereby significantly enhancing the computational capacity of multi-digit adders. It essentially applies the principle of chemical equilibrium, using a "divide and conquer" approach to enable the circuit to regulate reactant concentration ratios, thus optimizing signal transmission efficiency. Compared with existing technologies, this study has achieved several major breakthroughs. Specifically, we have achieved 17 consecutive carries in ternary form, which, to the best of our knowledge, represents the highest computational scale reported to date for DNA adders. Moreover, compared with several state-of-the-art DNA adders, our adder performs favorably across various evaluation metrics (e.g., scale per strand, strands per digit; see Supporting Information, Section S1 for detailed comparisons). Furthermore, biochemical experimental results confirm that our proposed CA strategy makes our DNA multi-digit ternary adder easily adaptable to calculations involving even higher digit counts. More importantly, the successful application of this new strategy, based on molecular competitive dynamics ($k$-control) and reactant concentration regulation (chemical equilibrium), provides novel insights and tools for designing and constructing more complex and robust chemical reaction-based molecular systems.

\section*{Results}
{\bf The Competitive Blocking Circuit.} The principle of the CB circuit is illustrated in Figure \ref{figure1}. It consists of two logic gates, denoted GATE$1_{ij}$ and GATE2, representing two levels of reaction. The CB circuit has two input signals, denoted as inputA$_{i}$ and inputB$_{j}$, along with a blocking signal B$_{in}$. The signal B$_{in}$ can be in a logic state of either $present$ or $absent$, indicating whether it participates in the reaction. Here, $i,j \in \{0,1,2\}$ represent the ternary numbers involved in the addition operation. When B$_{in}$ is in the $absent$ state, the CB circuit receives inputA$_{i}$ and inputB$_{j}$ as inputs, activating the first level of reaction. This means that the two inputs interact with GATE$1_{ij}$ to produce an output signal, denoted as SUM1. Conversely, when B$_{in}$ is $present$, it reacts with GATE$1_{ij}$ , effectively consuming GATE$1_{ij}$ and thereby inhibiting the first level of reaction between GATE$1_{ij}$ and the inputs inputA$_{i}$ and inputB$_{j}$. Furthermore, the presence of B$_{in}$ can activate the second level of the reaction, leading to a reaction with GATE$2$ and producing a different output signal, denoted as SUM2.

The CB mechanism can be expressed via the following reactions:


\begin{equation}\label{equa-1}
\mathrm{input A} _i + \mathrm{input B} _j + \mathrm{GATE1}_{ij} \xrightarrow{k_1} \mathrm{SUM}1 + \mathrm{waste}1
\end{equation}

\vspace{-0.6cm}
\begin{equation}\label{equa-2}
\mathrm{B_{in}} + \mathrm{GATE1}_{ij} \xrightarrow{k_2} \mathrm{waste}2
\end{equation}

\vspace{-0.6cm}
\begin{equation}\label{equa-3}
\mathrm{B_{in}} + \mathrm{GATE2} \xrightarrow{k_3} \mathrm{SUM}2 + \mathrm{waste}3
\end{equation}

Specifically, GATE1$_{ij}$  is a stable structure formed through base pairing in the $a_{i}$, $t_b$, and $b_{j}$ domains, while the toehold sites $t_{a}^*$ and $c^*$ remain exposed. The inputA$_{i}$ strand contains the toehold $t_a$ and the long domain $a_{i}$. The toehold $t_a$ of inputA$_{i}$ initiates a TMSD by pairing with the $t_a^*$ site on GATE1$_{ij}$ and continues base pairing until the strand incorporating $c^*$, $a_{i}$, and $t_b$ is displaced (The short domain $t_b$ will break free because the two longer domains $c^*$ and $a_i$ are in a free state). As a result, the $t_b^*$ site becomes exposed. Since inputB$_{j}$ contains the toehold $t_b$ and the long domain $b_{j}$, it undergoes a similar TMSD by pairing with the $t_b^*$ site, thus displacing the fluorescently labeled single strand $b_{j}$. This indicates the generation of SUM1. This process is described by Equation \ref{equa-1} and illustrated in Figure \ref{figure1}. 

\begin{figure*}[!t]
    \centering
    \includegraphics[width=1\linewidth]{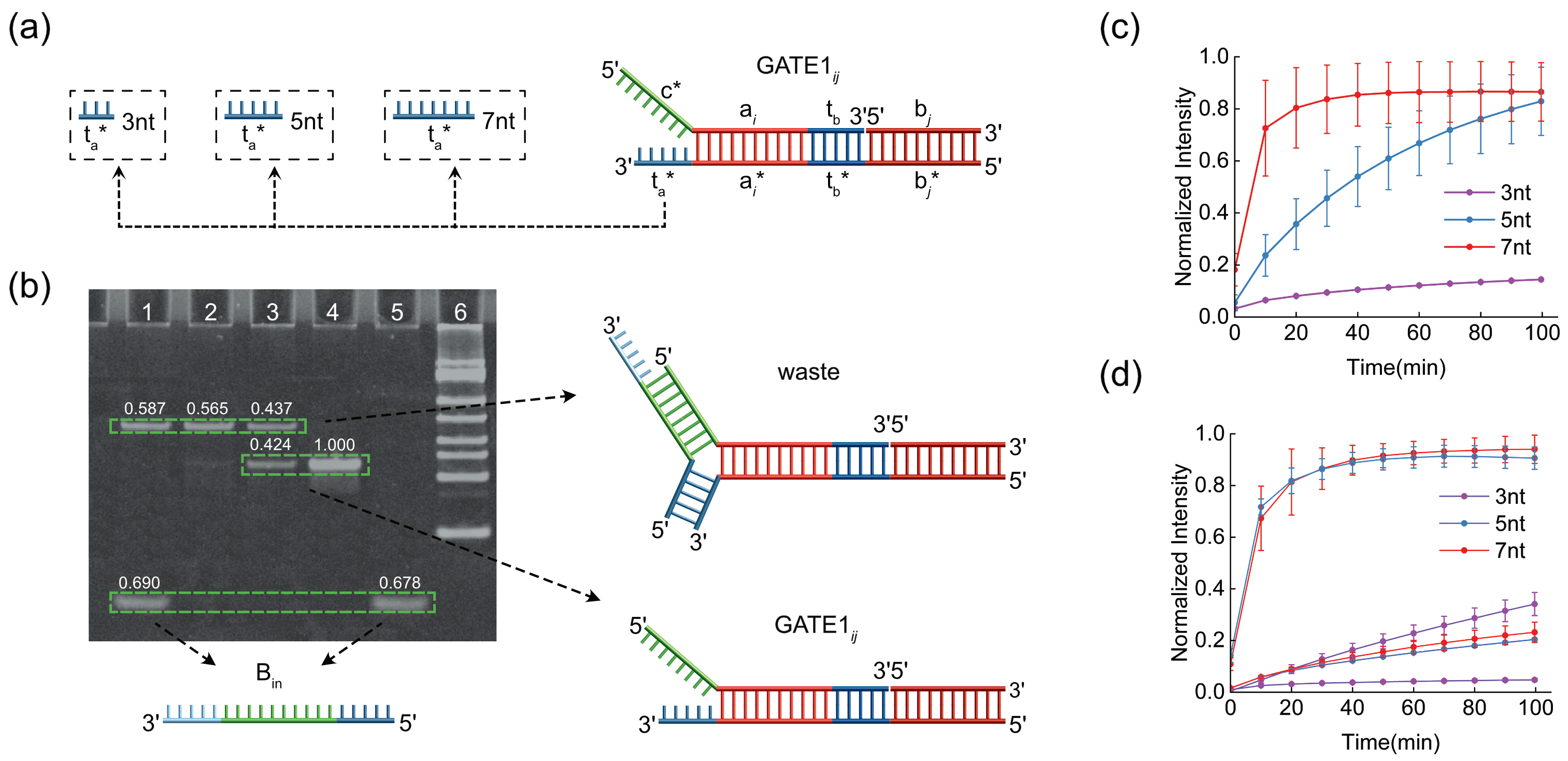}
   \caption{\small (a) Test schemes for different toehold lengths in the GATE1$_{ij}$ structure. (b) Polyacrylamide gel electrophoresis (PAGE) results show binding between B$_{in}$ and GATE1$_{ij}$ in the CB circuit at different concentration ratios, where the relative concentration of each band is determined by its brightness. (c) Fluorescence detection results for SUM1 product generated by the CB circuit when the toehold lengths of GATE1$_{ij}$ are 3nt, 5nt, and 7nt, respectively. (d) Fluorescence results for SUM2 product generated by the CB circuit when the toehold lengths of GATE1$_{ij}$ are 3nt, 5nt, and 7nt. Solid lines in this figure represent the fluorescence signal of SUM2, while dashed lines indicate the leakage of SUM1 due to B$_{in}$ not fully locking GATE1$_{ij}$.} 
    \label{figure2}
\end{figure*}

The blocking strand B$_{in}$  contains the toehold $t_a$ along with the long domains $c$ and $t_x$. The toehold $t_a$ binds to the $t_a^*$ site of GATE1$_{ij}$, effectively locking it through the complementary base pairing principle of DNA (Equation \ref{equa-2} and Figure \ref{figure1}). Given that the toehold $t_a$ comprises only five bases, there is a risk of detachment after locking solely by this interaction. To mitigate this risk, we have added domain $c^*$  to the 5’ end of domain $a_{i}$ to enhance the stability of B$_{in}$'s binding to GATE1$_{ij}$, ensuring that the toehold $t_{a}^*$ remains firmly locked. We conducted an evaluation of the binding stability of B$_{in}$ to GATE1$_{ij}$ both before and subsequent to the addition of domain $c^*$. The results indicate that GATE1$_{ij}$ exhibits a higher binding efficacy with B$_{in}$ following the incorporation of domain $c^*$ (see Supporting Information S4). This design effectively prevents the toehold $t_a$ in inputA$_{i}$ from engaging in a TMSD with the $t_a^*$ site of the gate structure, thereby ensuring that the output strand modified with fluorescence is not displaced. Upon locking GATE1$_{ij}$, the reaction proceeds with GATE$2$. GATE$2$ is a stable structure formed through base pairing in the $c$ domain, while the toehold site $t_{a}^*$ remains exposed. The blocking strand B$_{in}$ engages in a TMSD with the exposed toehold site $t_a^*$ of GATE$2$ (Equation \ref{equa-3} and Figure \ref{figure1}), leading to the displacement of the output strand that carries the fluorescence modification. The resulting computational output, SUM2, can be determined by analyzing the fluorescent signal.

The length of the toehold is pivotal in determining the rate of the TMSD \cite{zhang2009control,xu2025dna}. We evaluated the effect of three common toehold lengths of 3nt, 5nt, and 7nt on our reaction rates (Figure \ref{figure2}a). The fluorescence data (Figure \ref{figure2}c) indicate that with a 3nt toehold, the increase in fluorescence intensity is minimal, reflecting a low reaction efficiency. In contrast, extending the toehold length to 5nt and 7nt markedly accelerates the reaction rate, achieving a peak within 30 minutes. Therefore, we excluded the 3nt toehold from further consideration. Additionally, the binding efficiency of B$_{in}$ to GATE1$_{ij}$ affects the accuracy of the first-level reaction, as a portion of it still occurs even when B$_{in}$ is present. We call this a leakage reaction. To address this, we examined SUM1 leakage across different toehold lengths. The experimental findings are illustrated in Figure \ref{figure2}d. On the one hand, when using a 3nt toehold, the fluorescence of the leakage reaction producing SUM1 is minimal; however, SUM2 formation is too slow. On the other hand, although the reaction with a 7nt toehold is faster than that with a 5nt toehold, it results in greater leakage of the incorrect result SUM1. Consequently, we opted for a 5nt toehold length to minimize the generation of leakage reaction.

\begin{figure*}[!t]
    \centering
    \includegraphics[width=1\linewidth]{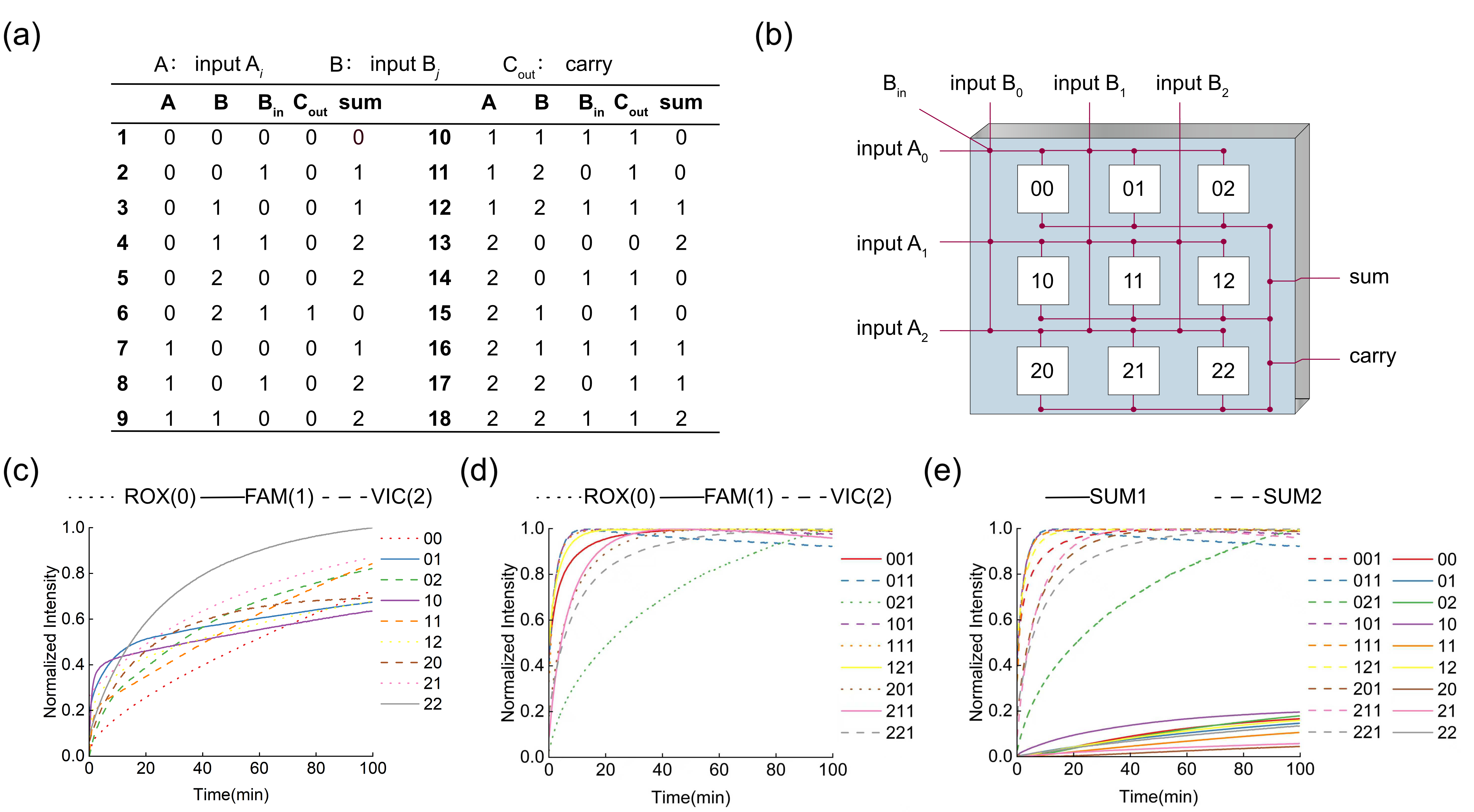}
    \caption{\small (a) Truth table for the ternary adder. (b) Modular processing of the ternary adder based on its truth table. (c) Fluorescence curves for all reactions of the half adder. The results were normalized to the maximum fluorescence across all samples, and a fluorescence reading above 0.5 was considered a valid output. (d) Fluorescence curves for the SUM2 result in the 1-trit full adder. The current results were normalized to their respective maximum fluorescence values, and the output reflects the fluorescence type with the highest fluorescence. (e) Fluorescence curves for all reactions of the 1-trit full adder. All fluorescence types in the current results were normalized to their respective maximum fluorescence values; solid lines represent SUM2. Dashed lines denote SUM1 leakage, attributed to the carry information from the preceding digit not completely locking the GATE1$_{ij}$.}
    \label{figure3}
\end{figure*}

Based on the experimental results for a toehold length of 5 nucleotides, as presented in Figure \ref{figure2}c and Figure \ref{figure2}d, the rates of critical reactions within the CB circuit are detailed as follows. The rate $k_1$ of Reaction 1 (Equation1, which is where inputA$_i$ and inputB$_j$ interact with GATE1$_{ij}$) is measured at $1.1876 \times 10^{-4}$ $\mu\text{M}$/s. In comparison, the rate $k_2$ of Reaction 2 (Equation2, which is where B$_{in}$ binds to GATE1$_{ij}$) is significantly higher at $1.8377 \times 10^{-3}$ $\mu\text{M}$/s, while the rate $k_3$ of Reaction 3 (Equation3, which is where B$_{in}$ interacts with GATE$2$) stands at $7.2992 \times 10^{-4}$ $\mu\text{M}$/s. Further details on the calculation of these rates are provided in Supporting Information S1.3. These results clearly indicate that the reaction rate of Reaction 2 is substantially greater than that of Reaction 1 and 3. This suggests that in the presence of B$_{in}$, it will preferentially and rapidly bind to GATE1$_{ij}$, effectively inhibiting Reaction 1. Following this, B$_{in}$ proceeds to react with GATE$2$. Conversely, when B$_{in}$ is absent, the CB circuit will proceed with Reaction 1. Additionally, to verify whether the B$_{in}$ successfully locks the gate, we examined the binding between B$_{in}$ and GATE1$_{ij}$ at varying B$_{in}$ concentrations.  
Polyacrylamide gel electrophoresis (PAGE) was performed; the results are shown in Fig.~\ref{figure2}b.  
From left to right, the molar ratios of B${in}$ to GATE1${ij}$ are 1.5:1, 1:1, 0.5:1, 0:1, and 1:0.  
Evidently, when B${in}$ is present, it binds efficiently to GATE1${ij}$.
In lane 3, a fraction of GATE1${ij}$ remains unbound because the B${in}$ concentration is lower than that of GATE1${ij}$. When the B${in}$ concentration is equal to or higher than that of GATE1${ij}$ (lanes 1 and 2), all GATE1${ij}$ molecules are bound by B$_{in}$. We therefore conclude that B$_{in}$ can effectively bind to GATE1$_{ij}$.

Moreover, we experimentally demonstrated the robustness of the CB circuit to sequence design choices and its tolerance to noise. First, we tested the stability of the CB circuit under different sequences. A total of nine distinct scenarios were examined, corresponding to the nine input–B$_{in}$ combinations of the ternary full adder (Figure S19). Subsequently, we tested the effect of B$_{in}$ on the CB circuit under varying GC content conditions (see Section S5 of the Supporting Information). The results indicate that the CB circuit consistently produces correct outputs across these varied sequences. Second, we evaluated the circuit's stability in the presence of noise, including mismatches and unrelated strands (see Section S5 of the Supporting Information). The findings confirm that the CB circuit remains functional and yields accurate results even under noisy conditions.

{\textbf{A Ternary DNA Full Adder Design  Using CB Circuit.} The design of the ternary full adder, based on the CB circuit, comprises four key components: converting input digits into DNA strands, computing the sum and carry, translating output signals into results, and extracting carry information as input to the next digit.  Each digit's input and output in a ternary adder can take three possible values (0, 1, or 2). This makes the conventional dual-rail strategy difficult to apply \cite{su2019high}. To address this, we employed a triple-rail method to represent the three input and output states. Specifically, we designate inputA$_{0}$, inputA$_{1}$, and inputA$_{2}$ to convey the inputs for 0, 1, and 2, while utilizing inputB$_{0}$, inputB$_{1}$, and inputB$_{2}$ for similar representations. Each inputA$_{i}$ consists of a toehold $t_a$ and domain $a_{i}$, and each inputB$_{j}$ comprises a toehold $t_b$ and domain $b_{j}$, where $i,j\in \{0,1,2\}$. Additionally, we employ three distinct fluorescent signals (ROX, FAM, VIC) to visualize the resulting outputs corresponding to 0, 1, and 2. It's worth noting that the fluorescent signal types assigned to each module's Gate$2$ and GATE1$_{ij}$ follow a specific pattern, as detailed in Table 4 of Section S15 in the Supporting Information.

Compared to traditional binary adders, the number of input combinations involved in this computation process differs significantly. The input combinations for a traditional binary adder amount to $2^3 = 8$ possibilities, as its inputs are limited to two states: 0 and 1. This simplicity enables straightforward implementation using conventional circuits. In contrast, the ternary adder presents a much more complex scenario; it features up to $2 \times 3^2 = 18$ input combinations (see Figure \ref{figure3}a), which poses significant challenges for implementation using standard circuits. To address this complexity, we modularize the design into nine distinct modules, termed \(ij\)-modules (where \(i,j \in \{0,1,2\}\)), inspired by the principles of computer memory addressing. Based on the CB circuit, these $ij$-modules are capable of computing $i+j+x$ (where $x$ is 0 or 1, representing the presence or absence of carry information), thereby generating the result information SUM1 or SUM2 (Figure \ref{figure3}b and Supporting Information, Figure S8 - S16). 

According to the truth table (Figure \ref{figure3}a), there are nine scenarios that result in carry information. We categorize these scenarios into two types for processing. 

The first type involves carry generation through a three-input AND gate. This includes cases 6, 10, and 14, where a carry is generated only in the presence of B$_{in}$ (the carry input from the previous digit). This indicates that the production of carry information relies on the inputs inputA$_{i}$, inputB$_{j}$, and B$_{in}$. Consequently, we implement this function using a three-input AND gate (GATE1$^{-}_{ij}$, where \( ij \in \{02, 11, 20\} \), as illustrated in Figure \ref{figure4}a). The structure of GATE1$^{-}_{ij}$ comprises four single strands, which stabilize through base pairing in the toehold regions \( t_a \) and \( t_b \), along with the longer domains \( a_{i} \), \( b_{j} \), and \( c \). At the same time, it reveals the toehold site \( t_a^* \) and the longer domain \( r \). When all three inputs, inputA$_{i}$, inputB$_{j}$, and B$_{in}$, are present, this gate produces the carry information strand \( cr \).
The specific reaction mechanism is depicted in Figure \ref{figure4}a. Initially, the toehold \( t_a \) of inputA$_{i}$ binds to the exposed toehold $t_{a}^*$ of GATE1$^{-}_{ij}$. This interaction facilitates the engagement of the long domain \( a_{i} \) of inputA$_{i}$ with the domain \( a_{i}^* \) within GATE1$^{-}_{ij}$, leading to the release of domain \( a_{i} \) from GATE1$^{-}_{ij}$. As \( t_a \) consists of only five bases, the strand \( a_i t_b \) will subsequently detach once domain \( a_{i} \) is liberated, thereby exposing the toehold \( t_b^* \). Subsequently, a parallel reaction will displace the strand \( b_{j} t_a \) upon the introduction of inputB$_{j}$, which in turn reveals the toehold \( t_a^* \). At this juncture, if the carry information B$_{in}$ from the previous digit is present, it will bind to the exposed toehold \( t_a^* \) and displace the information strand \( cr \) through a TMSD.

\begin{figure*}[!t]
    \centering
    \includegraphics[width=1\linewidth]{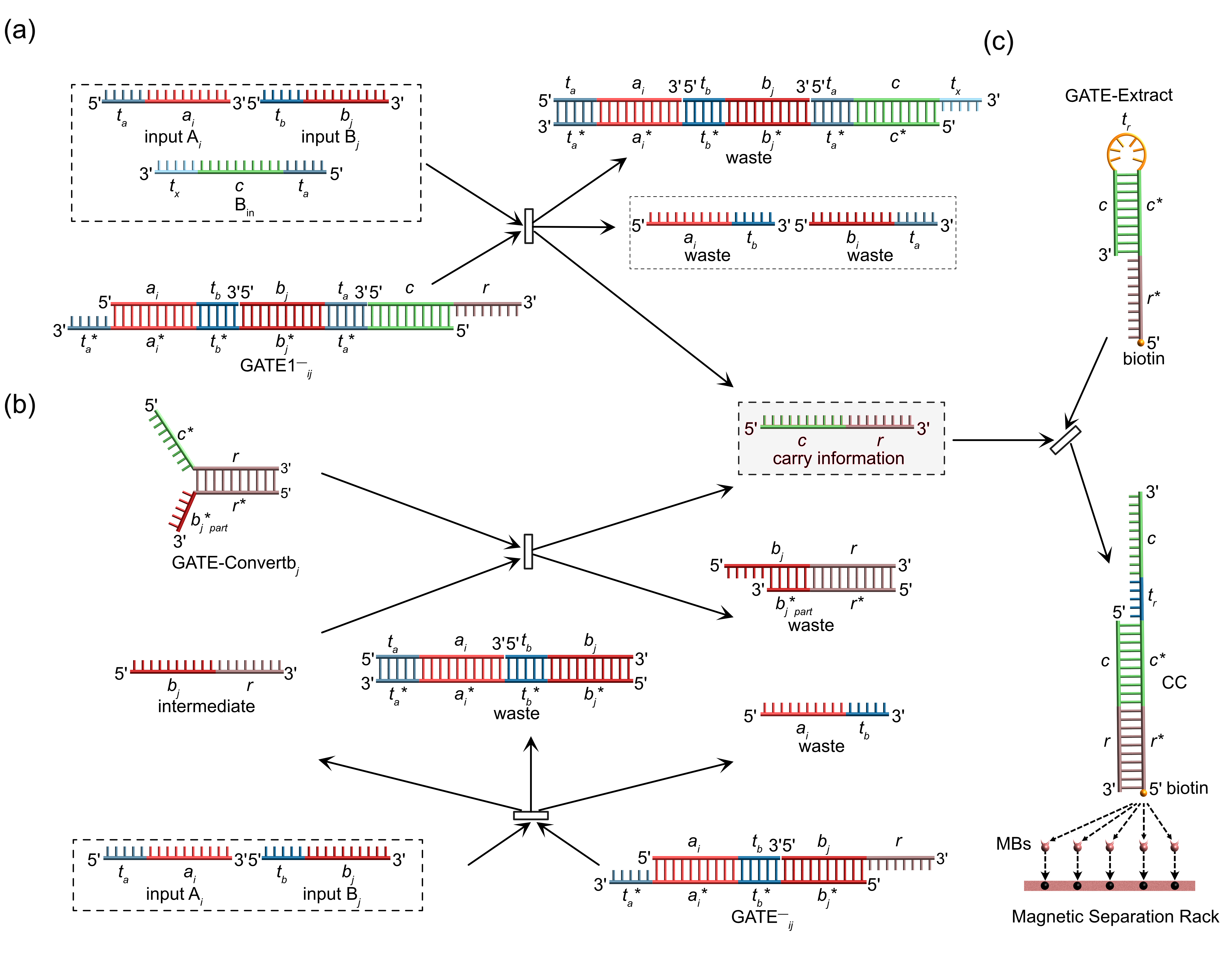}
    \caption{\small Generation and processing of carry information strands. (a) Schematic diagram of the three-input AND gate reaction for processing the first type of scenarios. (b) Schematic diagram of the two-input AND gate reaction for processing the second type of scenarios. (c) Extraction operation for the carry information strands.}
    
    \label{figure4}
\end{figure*}

The second type involves carry generation through a two-input AND gate in conjunction with a conversion gate. This category includes cases 11, 12, 15, 16, 17, and 18. In these scenarios, a carry is generated for the subsequent digit regardless of the carry input B$_{in}$. This indicates that the carry information relies exclusively on inputA$_{i}$ and inputB$_{j}$. We implement this function using a two-input AND gate GATE$^{-}_{ij}$, where \( ij \in \{12, 21, 22\} \) (see Figure \ref{figure4}b)). The structure of GATE$^{-}_{ij}$ consists of three single strands stabilized through base pairing at the toehold \( t_b \), along with the longer domains \( a_{i} \) and \( b_{j} \). It also features the toehold site \( t^*_a \) and a longer domain \( r \). When both inputA$_{i}$ and inputB$_{j}$ are present, this gate produces an output. The specific reaction mechanism is illustrated in Figure \ref{figure4}b. Similar to the first type, when inputA$_{i}$ and inputB$_{j}$ are both present, the strands \(a_i t_b\) and \( b_{j} r\) are sequentially displaced. It is important to highlight that this reaction does not directly yield the carry information strand \(cr\); instead, it produces an intermediate strand \( b_{j} r\). To resolve this issue, we devised a conversion gate GATE-Convertb$_j$ for \(j=1,2\) (as shown in Figure \ref{figure4}b). The structure of GATE-Convertb$_j$ consists of two single strands, one of which contains the necessary carry information. These strands are stabilized through base pairing involving their domain \(r\). Furthermore, the gate includes a toehold site \( b_{j}^* \) and a domain \(c\). When the intermediate strand \( b_{j} r\) is available, this gate will produce an output, effectively transforming \( b_{j} r\) into the carry information strand \(cr\) for subsequent extraction processes.

To extract the carry information and transfer it to the subsequent digit calculation, we designed an extractor referred to as GATE-Extract (Figure \ref{figure4}c). This structure forms a hairpin shape through self-complementary base pairing between the domains $c$ and $c^*$, with the toehold $t_r$ embedded within the hairpin. As shown in Figure \ref{figure4}c, when the carry information is present in the test tube, the domain $r^*$ of the hairpin engages in a TMSD with the domain $r$ of the information strand $cr$. This reaction opens the hairpin structure, revealing the toehold $t_r$ and the extended domain $c$. The GATE-Extract is modified with a biotin group at the 5' end. By leveraging the strong affinity between streptavidin MBs and biotin, GATE-Extract is anchored to the MB surface, thereby completing the extraction of the carried information.

{\textbf{An implementation of Ternary DNA Half-Adder and 1-trit Full Adder Using CB circuit.} A ternary half-adder is designed to add two ternary digits, producing two outputs: a sum \( S_{out} \) and a carry \(C_{out}\). This operation is performed without considering carry information. To calculate the sum of two inputs, inputA$_{i}$ and inputB$_{j}$, we utilize the \( ij \)-module. Let’s examine the case where \(i=2\) and \(j=2\). Notably, other cases follow a similar method. In our implementation, inputA$_{2}$ and inputB$_{2}$ are introduced into the \( ij \)-module. These inputs interact with GATE1$_{22}$, resulting in the gradual displacement of the \( b_{2} \) strand from GATE1$_{22}$. Since the 3' end of strand \(b_j\) is modified with a FAM fluorescent group, and the bottom strand of GATE1$_{22}$ is equipped with a BHQ1 quencher group, the FAM signal is triggered upon the displacement of strand \(b_2\). This results in an \(S_{out}\) value of 1, as illustrated in Figure \ref{figure3}c, curve 22.  Concurrently, inputA$_{2}$ and inputB$_{2}$ in the test tube participate in a TMSD with GATE$^{-}_{ij}$ (Figure \ref{figure4}b), which displaces the strand \( b_{j}\)\(r\). Subsequently, the strand \( b_{j}\)\(r\) interacts with  GATE-Convertb$_{j}$ to produce the carry \( C_{out} \) strand. In addition to the experiment described above, we also tested the other eight input combinations using \( ij \)-modules for \( i,j\in \{0,1,2\} \), where \( ij \neq 22 \). The fluorescence output of each module matches the expected result (Figure \ref{figure3}c and Supporting Information, Figure S17), further validating the reliability of the ternary half-adder design.

We constructed a 1-trit full adder using our CB circuit. In comparison to the half-adder, the full adder includes an additional carry input, B$_{in}$, from the previous digit, making it a third input. This circuit also produces two outputs: a sum \(S_{out}\) and a carry \(C_{out}\). We employed the \(ij\)-module to compute the addition of two addend inputs alongside the carry input B$_{in}$ (serving as the blocking strand). For illustrative purposes, we set \(i=2\) and \(j=2\); other cases can be treated similarly. The implementation involves four logic gates: GATE1$_{22}$, GATE2, GATE$^{-}_{22}$, and GATE-Convertb$_2$.  Specifically, the CB circuit allows B$_{in}$ to block the toehold domain of GATE1$_{22}$ (Equation \ref{equa-2}), preventing it from reacting with inputA$_2$ and inputB$_2$. At the same time, B$_{in}$ engages in strand displacement with GATE2 (Equation \ref{equa-3}), resulting in the release of the strand \(c\) labeled with a VIC fluorophore at its 3' end. The detection of this VIC fluorescence signal indicates that \(S_{out}=2\). Concurrently, inputA$_2$ and inputB$_2$ undergo strand displacement with GATE$^{-}_{22}$ (Figure \ref{figure4}b), releasing the strand \(b_2r\), which then interacts with GATE-Convertb$_2$ to ultimately generate the carry \(C_{out}\). Experimental results show that B$_{in}$ effectively blocked GATE$1_{22}$, limiting the fluorescence intensity of the leakage reaction between inputA$_2$ and inputB$_2$ with GATE1$_{22}$ to below 0.5 (Figure \ref{figure3}e). This mechanism, which redirects the reaction to GATE2 and B$_{in}$, yielded VIC fluorescence output corresponding to the correct result 2 (Supporting Information, Figure S18). In addition to the aforementioned experiments, we also evaluated the other eight input combinations using $ij$-modules (where $i,j \in \{0,1,2\}$ and $ij \neq 22$). Each of these combinations successfully generated a fluorescence signal indicating the correct output (Figure \ref{figure3}d and Supporting Information, Figure S18). Furthermore, in contrast to the half-adder (where outputs were exclusively SUM1 result information), when B$_{in}$ is present, all input combinations yielded SUM2 result information. Crucially, the corresponding SUM1 result information (generated by the leakage reaction) consistently remained below half of the SUM2 result information (Figure \ref{figure3}e). To further analyze the absolute output and absolute leakage for each input combination in the presence of B$_{in}$, we normalized all fluorescence signals (including both valid outputs and leakage) to the highest fluorescence value. The experimental results show that the normalized SUM1 leakage signal intensity remains consistently below half of the SUM2 output signal intensity. (see Section S9 of the Supporting Information) This decisively demonstrates the effectiveness of the CB circuit in the adder.

{\textbf{The Implementation of Multi-digit Ternary DNA Full Adder Using CB Circuit.} To effectively manage the propagation of carry information for each digit in a multi-trit adder, we utilized streptavidin MBs to extract and transfer carry information. Streptavidin is a protein known for its exceptional binding affinity to biotin \cite{ozawa2017role,zhang2015reconfigurable}. Since the carry information strand is not the B$_{in}$ strand required for subsequent carry propagation, we designed an amplification-and-conversion reaction (Figure \ref{figure5}a). First, the GATE-amplifier transforms the Carry Carrier (CC) into B$_{in}$ strands. Then, to compensate for signal loss inherent in TMSD and for losses during magnetic-bead extraction and transfer, we introduce fuel strands that further react with the intermediate strand generated from CC and GATE-amplifier, regenerating CC and thereby cyclically producing more B$_{in}$ strands for downstream carry calculations. Experiments show that this amplification reaction increases B$_{in}$ strand yield by approximately threefold (see Section S10 of the Supporting Information). The specific procedure is detailed below: The completed computational system is positioned on a magnetic rack, where GATE-Extract bound to streptavidin MBs is attracted to the bottom of the test tube via magnetism. After removing the supernatant (which contains unneeded waste), the GATE-amplifier and fuel strands are introduced. During experiments, it was observed that even in the absence of CC, a fluorescent signal was still produced. The reason this happened is that the GATE-amplifier, intended for signal amplification, can react similarly to B$_{in}$, though with lower efficiency. To mitigate this issue, we modified the 5' end of the single strand located below the GATE-amplifier with biotin. After signal amplification, we placed the reaction system on a magnetic rack again, thereby drawing the GATE-amplifier to the bottom of the test tube. Following this, the supernatant was collected as the B$_{in}$ input for the next digit. This approach prevents misrecognition and ensures the accurate transfer of carry information.

\begin{figure*}[!t]
    \centering
    \includegraphics[width=1\linewidth]{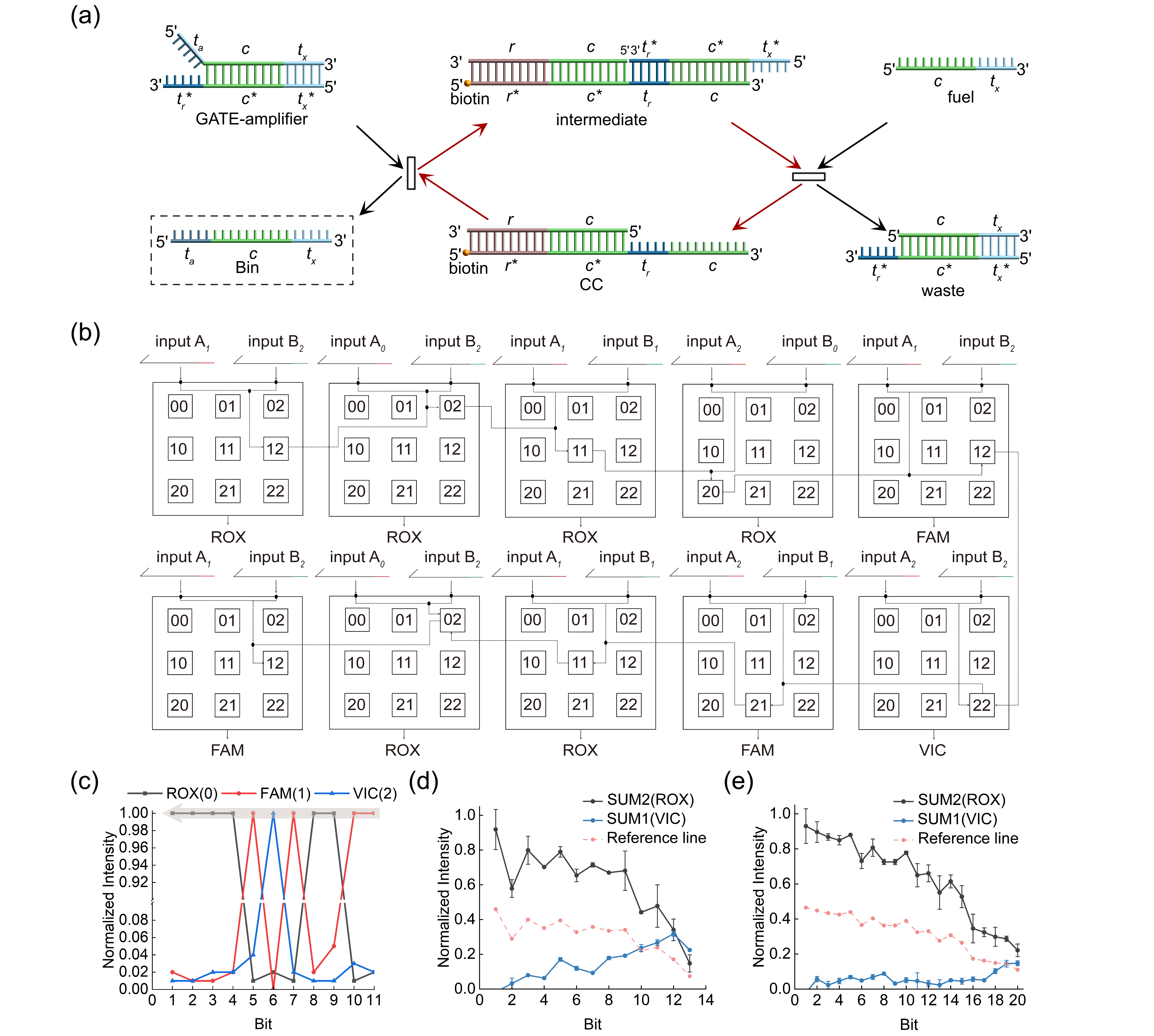}
    \caption{\small (a) Schematic diagram of the catalytic reaction for the carry information strand. (b)  Reaction network of a multi-trit full adder calculating 1012122101 + 2210221122. (c) The sum of two ten-digit ternary numbers. The final sum is read from the high digit to the low digit. (d) Computational performance of the ternary adder under continuous-carry conditions without the CA strategy, illustrated with the example 102102102101 + 120120120122, where the reference line represents half of the SUM2 signal, i.e., the leakage threshold. (e)  Computational performance of the ternary adder under continuous carry conditions with the CA strategy, illustrated with the example: 2102102102102102101 + 0120120120120120122, where the reference line represents half of the SUM2 signal, i.e., the leakage threshold.}
    
    \label{figure5}
\end{figure*}

To evaluate the feasibility of the multi-digit ternary full adder, we conducted an operational test using a ten-digit ternary full adder, focusing on the addition $1012212101 + 2211220122$. The procedure began by adding inputA$_{1}$ and inputB$_{2}$ to the designated test tube for the 12-module, followed by measurement of the resulting digit's fluorescent signal. Subsequently, carry information was extracted through magnetic selection and subjected to a signal amplification reaction. After another round of magnetic selection, the supernatant was transferred to the next digit calculation module along with the next digit's inputs (inputA$_0$ and inputB$_2$). The output from the fluorescent reporter gate was then recorded. This process continued until all digit calculations were complete (Figure \ref{figure5}b). Additionally, to enable an intuitive comparison of leakage rates at each digit, we predetermined the fluorescence intensity ratios among the three dyes through 15 independent experiments (see Section S11 of the Supporting Information) and uniformly normalized all subsequent data accordingly. The results are illustrated in Figure \ref{figure5}c, where the green, orange, and blue curves correspond to the fluorescence signals of ROX, FAM, and VIC, representing values of 0, 1, and 2, respectively. For each digit, the number indicated by the fluorescence species with the maximum fluorescence value (where all other fluorescence readings fall below half of this maximum) is deemed the final result for that digit. The overall result presented in the figure is $11001210000$, thereby confirming both the feasibility and accuracy of the design.

To rigorously evaluate the scalability of the CB-circuit-based ternary adder, we adopted ``consecutive carry capability'' as the key metric. Should consecutive carries terminate, the B$_{in}$ strand concentration is reset, and the true scalability is masked; hence, the upper limit of consecutive carries directly determines the scalability boundary. We thus designed a full consecutive-carry test: the adder achieves 100\,\% success for up to eight consecutive carries, whereas success drops to 33.33\,\% when consecutive carries reach the 9th and 10th digits (Figure 5d). Restricting our analysis to 100\,\% successful cases, we conclude that the consecutive-carry capability of the CB ternary adder is eight digits.

{\textbf{Experimental Expansion of Digit Positions for the Multi-digit DNA Ternary Full Adder.} Due to the inherent limitations of biochemical experiments, multi-digit DNA adders face a core challenge during actual operation: as the number of consecutive carry digits increases, the quantity of the carry information strand progressively diminishes, leading to computational errors. This is the fundamental reason impacting the scalability of DNA adders. Our design addresses this challenge by decoupling the computation of the result digit from that of the carry. The CA strategy allows us to enhance the performance by adjusting the concentration ratios of the relevant components. In particular, reducing the concentration of GATE$1_{ij}$ enables it to be more effectively bound by B$_{in}$ (the carry generated from the previous digit), thereby minimizing signal leakage.

To validate this finding, we compared the experimental results from two different concentration configurations: 1) a GATE$^{-}_{ij}$ concentration of 1 \(\mu\)M and a GATE$1_{ij}$ concentration of 0.4 \(\mu\)M; and 2) a GATE$^{-}_{ij}$ concentration of 1 \(\mu\)M and a GATE$1_{ij}$ concentration of 0.2 \(\mu\)M. The results, illustrated in Figure \ref{figure5}d, indicate that with the higher concentration of GATE$1_{ij}$ in the first case, the carry information from the preceding digit must block more GATE$1_{ij}$ molecules, leading to a decrease in the amount of carry information available for carry calculation. Consequently, as the number of digits increases, the carry information exhibits a trend of accelerated attenuation, which in turn results in a significant reduction in the carry information contributing to the sum calculation. The fluorescence signal for the sum digit decays notably faster in this scenario compared to the second case. Note that even without the introduction of the CA strategy, we could still achieve eight consecutive carry calculations (Figure \ref{figure5}d). This is because our judgment was based on the maximum fluorescence value for each digit, requiring other fluorescence values to be less than half that maximum. Therefore, even as the information continuously decayed, the leakage rate(i.e., the ratio of the maximum fluorescence value corresponding to non-maximal fluorescence types to the maximum fluorescence value) remained below 0.5, allowing us to perform consecutive multi-digit carry calculations. Based on this, we further evaluated changes in the leakage rate as the number of digits increased following the introduction of the CA strategy. The results show that the leakage rate was significantly reduced after the introduction of the CA strategy (Figure \ref{figure5}d and Figure \ref{figure5}e). Additionally, the CA strategy helps to reduce signal attenuation. In particular, as the digit count increases, the adder employing the CA strategy demonstrates a slower decay rate in the fluorescence signal that indicates the correct result (see Figure \ref{figure5}d) compared to the adder that does not utilize the CA strategy (see Figure \ref{figure5}e). This strongly demonstrates the effectiveness of our proposed CA strategy.

To assess the practical upper limit of the CA strategy, we performed a 17-consecutive-carry ternary addition and successfully obtained the correct output. However, the 18th carry failed because the leakage rate exceeded the detection threshold, thereby establishing the current system's maximum reliable depth at 17 carries (see Figure 5e). Additionally, we examine in detail the applicability of the CA strategy in Section S13 of the Supporting Information. This achievement represents a significant advancement over the current state-of-the-art multi-bit adder based on DNA circuits, which has only achieved four consecutive carries. In terms of computational scale, our result is $8.61 \times 10^6$ times that of the recently reported study (detailed calculations are provided in Supporting Information s1.2); compared to an advanced binary adder implemented using non-DNA circuit technology (which achieves 25 consecutive carries), our result is 3.85 times its computational scale. Furthermore, given differences in the number of strands used across adders, we introduced the "scale/strand" metric to more reasonably evaluate the computational capability of the adders. Under this metric, our achievement reaches $2.41 \times 10^6$ times and 30.6 times that of the two aforementioned comparison targets, respectively. Additionally, we analyzed the per-digit leakage of our adder, all of which remained within acceptable ranges (detailed calculations are provided in Supporting Information s1.1). Furthermore, our adder achieves the best performance in terms of the number of strands per digit (detailed calculations are provided in Supporting Information s1.2). This finding provides a promising direction for future design optimization of higher-digit full adders.

To intuitively illustrate the scalability advantage of the higher radix, we first developed 9-digit binary and 8-digit ternary full adders (both featuring consecutive carries) within the same CB-circuit core architecture. As depicted in Figure 5d and Figure S24, both types operate with a comparable digit count, yet the computational capability of the binary adder is only about 7.79\% that of the ternary adder (refer to Supporting Information S12 for further details). Additionally, using the CA strategy, we constructed 17-digit binary and 17-digit ternary full adders (with consecutive carries). The experimental outcomes presented in Figure 5e and Figure S25 demonstrate that both adders once again maintain the same digit count; however, the computational scale of the binary adder is a mere 0.1\% of that of the ternary adder (see Supporting Information S12 for more details). These two adders achieve comparable digit depths, as the per-digit signal attenuation is similarly aligned under the identical CB/CA core architecture (as illustrated in Figures S26 and S27). Thus, within the same framework, the ternary implementation clearly offers a significant advantage.


\section*{CONCLUSIONS}
This paper reports the first experimental realization of a DNA‑based ternary full adder. To achieve this, we designed a chemically kinetic logic control circuit (CB circuit) that cleverly exploits differences in reaction rate constants to enable dynamic selection and blocking of reaction pathways. Second, by applying the principles of chemical equilibrium, we propose a CA strategy based on a divide-and-conquer approach. This strategy empowers the circuit to regulate reactant concentration ratios, thereby optimizing signal transmission efficiency. Subsequently, these two innovations have led to the successful realization of a ternary DNA full adder. This innovative ternary DNA adder significantly overcomes the computational scale limitations of traditional DNA adders. Experimental results demonstrate that the maximum computational scale achievable by the DNA adder with consecutive carries is improved by approximately $8.61 \times 10^6$ times compared to existing DNA circuit-based four-consecutive-carry adders. Meanwhile, it also achieves a 3.85-fold improvement in computational scale over advanced binary adders implemented using thermodynamic methods, representing a significant enhancement. Moreover, our adder also performs favorably across various other evaluation metrics (e.g., scale per strand, strands per digit; see Supporting Information, Section S1 for detailed comparisons). While achieving infinite carry propagation remains a challenge, the experimental findings fully demonstrate the validity and efficiency of our approach.

Although we have not yet directly characterized addition circuits beyond seventeen consecutive carries, several strategies can further extend scalability: (i) introducing additional leakage-filter modules or lower-leakage strand-displacement schemes (e.g., using clamps in the seesaw circuits); (ii) fine-tuning the CA strategy to further strengthen the carry-information strand; (iii) circuit optimization, such as employing enzymes to achieve more efficient strand displacement and thus reduce signal attenuation; and (iv) spatial localization, anchoring computational modules at distinct locations (e.g., specific sites on DNA origami or on separate magnetic beads/nanoparticle surfaces) to prevent signal molecules from diffusing throughout the bulk solution, minimize crosstalk, and ensure that signals are delivered to neighboring modules along predefined paths. Collectively, these approaches provide a clear roadmap for pushing the DNA ternary-adder architecture beyond the current experimentally demonstrated number of carry digits.

The realization of a ternary DNA full adder represents a significant advancement in biomolecular computing. Our research not only enhances computational capabilities but also helps reduce costs, making DNA computing more practical and sustainable. Additionally, the CB circuit can process multiple input signals based on their priority, providing a powerful tool for developing more complex DNA computing systems. This also highlights the immense potential of DNA computing in advancing the chemistry of molecular information processing. Moreover, the CA strategy and the module partitioning method provide a new perspective for reducing the impact of leakage reactions on circuit performance. More importantly, the successful application of the CB circuit and CA strategy provides new insights and tools for designing and constructing more complex and robust chemical reaction-based molecular systems, extending beyond computation into areas such as sensing, diagnostics, and drug-delivery control.

To develop a more intelligent, automated ternary multi-digit full adder, we aim to integrate microfluidic technology with other techniques. This integration will enable the immobilization of magnetic beads within microchannels and achieve fully automated, continuous addition operations through programmed valve‑based flow switching. The ternary full adder has the potential to optimize the performance of complex systems and may offer unique advantages in areas such as artificial intelligence, neural networks, and cryptography in the future.

\section*{Methods}
{\textbf{Materials.} DNA oligonucleotides were purchased from Sangon Biotech (Shanghai) Co., Ltd. and purified by high-performance liquid chromatography (HPLC). All DNA strand sequences used in biochemical experiments were initially designed using NUPACK and subsequently manually optimized to ensure GC content remained between 30\% and 60\%. The full list of sequences used in this study is provided in Table 3 of Section S14 in the Supporting Information. Unmodified DNA strands were dissolved in $1\times$ TE buffer and stored at $-20\,^\circ\text{C}$. Fluorescently labeled or quencher-modified strands were dissolved in deionized water and stored at $-20\,^\circ\text{C}$ in the dark. DNA concentrations were measured using a NanoPhotometer N120. Prior to experiments, DNA samples were prepared by mixing with 12.5 mM $MgCl_2$ in $1\times$ TE buffer.

The DNA complexes were assembled by mixing equimolar amounts of the corresponding single-stranded DNA oligonucleotides and then annealing in a PCR thermocycler. The annealing protocol consisted of an initial denaturation at $95\,^\circ\text{C}$ for 2 min, followed by a gradual cooling: from $95\,^\circ\text{C}$ to $65\,^\circ\text{C}$ at a rate of $0.1\,^\circ\text{C}$ every 3s, then to $35\,^\circ\text{C}$ at $0.1\,^\circ\text{C}$ every 6s, and finally to $4\,^\circ\text{C}$ at $0.1\,^\circ\text{C}$ every 3s. The annealed complexes were stored at $4\,^\circ\text{C}$ for subsequent use. For fluorescently labeled or quencher-modified complexes, the cooling rate after the initial denaturation ($95\,^\circ\text{C}$, 2 min) was set to $0.1\,^\circ\text{C}$ every 6s down to $4\,^\circ\text{C}$, after which the samples were stored at $4\,^\circ\text{C}$ until use.

{\textbf{Fluorescence Kinetics Experiments.} In fluorescence-kinetics assays, the target DNA concentration was inferred from the measured fluorescence intensity. DNA samples were prepared according to the experimental design, with 1$\times$ defined as 1 $\mu\text{M}$. Final concentrations corresponded to 2$\times$ input gate and 1$\times$ reaction gate. Fluorescence was monitored using a QuantStudio 3\&5 Real-Time PCR System (Thermo Fisher Scientific, Waltham, MA, USA) equipped with a 96-well fluorescence plate reader. The thermal profile was set as follows: during the holding stage, the temperature was decreased to $1.6\,^\circ\text{C}$ at $4\,^\circ\text{C}$/s and held for 10s before the PCR stage; subsequently, the temperature was ramped up to $23\,^\circ\text{C}$ at $3\,^\circ\text{C}$/s, and fluorescence readings were acquired every 10s. All fluorescence curves were normalized to allow for direct comparison across experiments.

{\textbf{Statistical Analysis.} Normalization method: Fluorescence correction was performed via 15 independent experiments---under identical circuit and sequence conditions, single-dye reactions (ROX, FAM or VIC) were run in full, the intensity of each dye was recorded in all three channels, and the brightest ROX was set as reference (1.00) to yield the correction ratio \text{ROX}:\text{FAM}:\text{VIC} = 1.00:0.939:0.261; all subsequent readings were corrected by these factors. Normalization formula: $x_{\text{norm}}=(x-x_{\min})/(x_{\max}-x_{\min})$. Data presentation: Results are shown as mean $\pm$ SD; consecutive-carry experiments used two replicates, whereas data from the half-adder and one-digit full adder are plotted using a single dataset; all remaining experiments used three replicates. Statistical analyses were performed using OriginPro 2024.

\bibliography{MSP-template}

@article{jeong2019tunnelling,
  title={Tunnelling-based ternary metal--oxide--semiconductor technology},
  author={Jeong, Jae Won and Choi, Young-Eun and Kim, Woo-Seok and Park, Jee-Ho and Kim, Sunmean and Shin, Sunhae and Lee, Kyuho and Chang, Jiwon and Kim, Seong-Jin and Kim, Kyung Rok},
  journal={Nature Electronics},
  volume={2},
  number={7},
  pages={307--312},
  year={2019},
  publisher={Nature Publishing Group UK London}
}

@article{hayes2001third,
  title={Third base},
  author={Hayes, Brian},
  journal={American scientist},
  volume={89},
  number={6},
  pages={490--494},
  year={2001}
}

@article{aguirre2010cmos,
  title={CMOS full-adders for energy-efficient arithmetic applications},
  author={Aguirre-Hernandez, Mariano and Linares-Aranda, Monico},
  journal={IEEE transactions on very large scale integration (VLSI) systems},
  volume={19},
  number={4},
  pages={718--721},
  year={2010},
  publisher={IEEE}
}

@article{elbaz2010dna,
  title={DNA computing circuits using libraries of DNAzyme subunits},
  author={Elbaz, Johann and Lioubashevski, Oleg and Wang, Fuan and Remacle, Fran{\c{c}}oise and Levine, Raphael D and Willner, Itamar},
  journal={Nature nanotechnology},
  volume={5},
  number={6},
  pages={417--422},
  year={2010},
  publisher={Nature Publishing Group UK London}
}

@article{orbach2014full,
  title={A full-adder based on reconfigurable DNA-hairpin inputs and DNAzyme computing modules},
  author={Orbach, Ron and Wang, Fuan and Lioubashevski, Oleg and Levine, Rapha{\"e}l David and Remacle, Francoise and Willner, Itamar},
  journal={Chemical Science},
  volume={5},
  number={9},
  pages={3381--3387},
  year={2014},
  publisher={Royal Society of Chemistry}
}

@article{lv2023dna,
  title={DNA-based programmable gate arrays for general-purpose DNA computing},
  author={Lv, Hui and Xie, Nuli and Li, Mingqiang and Dong, Mingkai and Sun, Chenyun and Zhang, Qian and Zhao, Lei and Li, Jiang and Zuo, Xiaolei and Chen, Haibo and others},
  journal={Nature},
  volume={622},
  number={7982},
  pages={292--300},
  year={2023},
  publisher={Nature Publishing Group UK London}
}

@article{mehra20132,
  title={2-bit comparator using different logic style of full adder},
  author={Mehra, Vandana Choudhary Rajesh},
  journal={Int. J. Soft Comput. Eng. IJSCE},
  volume={3},
  pages={277--9},
  year={2013}
}

@article{imana2017fast,
  title={Fast bit-parallel binary multipliers based on type-I pentanomials},
  author={Imana, Jose L},
  journal={IEEE Transactions on Computers},
  volume={67},
  number={6},
  pages={898--904},
  year={2017},
  publisher={IEEE}
}

@article{navi2009two,
  title={Two new low-power full adders based on majority-not gates},
  author={Navi, Keivan and Moaiyeri, Mohammad Hossein and Mirzaee, Reza Faghih and Hashemipour, Omid and Nezhad, Babak Mazloom},
  journal={Microelectronics journal},
  volume={40},
  number={1},
  pages={126--130},
  year={2009},
  publisher={Elsevier}
}

@article{kouretas2012low,
  title={Low-power logarithmic number system addition/subtraction and their impact on digital filters},
  author={Kouretas, Ioannis and Basetas, Charalambos and Paliouras, Vassilis},
  journal={IEEE Transactions on Computers},
  volume={62},
  number={11},
  pages={2196--2209},
  year={2012},
  publisher={IEEE}
}

@article{seelig2006enzyme,
  title={Enzyme-free nucleic acid logic circuits},
  author={Seelig, Georg and Soloveichik, David and Zhang, David Yu and Winfree, Erik},
  journal={science},
  volume={314},
  number={5805},
  pages={1585--1588},
  year={2006},
  publisher={American Association for the Advancement of Science}
}

@article{turberfield2003dna,
  title={DNA fuel for free-running nanomachines},
  author={Turberfield, Andrew J and Mitchell, JC and Yurke, Bernard and Mills Jr, Allen P and Blakey, MI and Simmel, Friedrich C},
  journal={Physical review letters},
  volume={90},
  number={11},
  pages={118102},
  year={2003},
  publisher={APS}
}

@article{zhang2009control,
  title={Control of DNA strand displacement kinetics using toehold exchange},
  author={Zhang, David Yu and Winfree, Erik},
  journal={Journal of the American Chemical Society},
  volume={131},
  number={47},
  pages={17303--17314},
  year={2009},
  publisher={ACS Publications}
}

@article{zhang2007engineering,
  title={Engineering entropy-driven reactions and networks catalyzed by DNA},
  author={Zhang, David Yu and Turberfield, Andrew J and Yurke, Bernard and Winfree, Erik},
  journal={Science},
  volume={318},
  number={5853},
  pages={1121--1125},
  year={2007},
  publisher={American Association for the Advancement of Science}
}

@article{weng2022cooperative,
  title={Cooperative branch migration: a mechanism for flexible control of DNA strand displacement},
  author={Weng, Zhi and Yu, Hongyan and Luo, Wang and Guo, Yongcan and Liu, Qian and Zhang, Li and Zhang, Zhang and Wang, Ting and Dai, Ling and Zhou, Xi and others},
  journal={ACS nano},
  volume={16},
  number={2},
  pages={3135--3144},
  year={2022},
  publisher={ACS Publications}
}

@article{wu2024plug,
  title={Plug-and-play module for reversible and continuous control of DNA strand displacement kinetics},
  author={Wu, Lang and Wang, Guan A and Li, Feng},
  journal={Journal of the American Chemical Society},
  volume={146},
  number={10},
  pages={6516--6521},
  year={2024},
  publisher={ACS Publications}
}

@article{liu2020cross,
  title={Cross-Inhibitor: a time-sensitive molecular circuit based on DNA strand displacement},
  author={Liu, Chanjuan and Liu, Yuan and Zhu, Enqiang and Zhang, Qiang and Wei, Xiaopeng and Wang, Bin},
  journal={Nucleic acids research},
  volume={48},
  number={19},
  pages={10691--10701},
  year={2020},
  publisher={Oxford University Press}
}

@article{tandon2020demonstration,
  title={Demonstration of arithmetic calculations by DNA tile-based algorithmic self-assembly},
  author={Tandon, Anshula and Song, Yongwoo and Mitta, Sekhar Babu and Yoo, Sanghyun and Park, Suyoun and Lee, Sungjin and Raza, Muhammad Tayyab and Ha, Tai Hwan and Park, Sung Ha},
  journal={ACS nano},
  volume={14},
  number={5},
  pages={5260--5267},
  year={2020},
  publisher={ACS Publications}
}

@article{huang2019versatile,
  title={Versatile and Homogeneous DNA Tetraplex Platform for Constructing Label-Free Logic Devices: From Design to Application},
  author={Huang, Dan and Yang, Chunrong and Yao, Ye and Li, Jicheng and Guo, Chen and Chen, Jianchi and Zhang, Yi and Yang, Shu and Yang, Qianfan and Tang, Yalin},
  journal={Chemistry--A European Journal},
  volume={25},
  number={28},
  pages={6996--7003},
  year={2019},
  publisher={Wiley Online Library}
}

@article{tang2025localized,
  title={A Localized Scalable DNA Logic Circuit System Based on the DNA Origami Surface},
  author={Tang, Zhen and Li, Shiyin and Chen, Chunlin and Zhou, Zhaohua and Yin, Zhixiang},
  journal={International Journal of Molecular Sciences},
  volume={26},
  number={5},
  pages={2043},
  year={2025},
  publisher={MDPI}
}

@article{adleman1994molecular,
  title={Molecular computation of solutions to combinatorial problems},
  author={Adleman, Leonard M},
  journal={science},
  volume={266},
  number={5187},
  pages={1021--1024},
  year={1994},
  publisher={American Association for the Advancement of Science}
}

@article{fan2020propelling,
  title={Propelling DNA computing with materials’ power: recent advancements in innovative DNA logic computing systems and smart bio-applications},
  author={Fan, Daoqing and Wang, Juan and Wang, Erkang and Dong, Shaojun},
  journal={Advanced Science},
  volume={7},
  number={24},
  pages={2001766},
  year={2020},
  publisher={Wiley Online Library}
}

@article{kim2018detection,
  title={The detection of a mismatched DNA by using hairpin DNA-templated silver nanoclusters},
  author={Kim, Seyeon and Gang, Jongback},
  journal={Analytical Biochemistry},
  volume={549},
  pages={171--173},
  year={2018},
  publisher={Elsevier}
}

@article{gasser2017sensing,
  title={Sensing of dangerous DNA},
  author={Gasser, Stephan and Zhang, Wendy YL and Tan, Nikki Yi Jie and Tripathi, Shubhita and Suter, Manuel A and Chew, Zhi Huan and Khatoo, Muznah and Ngeow, Joanne and Cheung, Florence SG},
  journal={Mechanisms of ageing and development},
  volume={165},
  pages={33--46},
  year={2017},
  publisher={Elsevier}
}

@article{spencer2021binding,
  title={The binding of monoclonal and polyclonal anti-Z-DNA antibodies to DNA of various species origin},
  author={Spencer, Diane M and Reyna, Angel Garza and Pisetsky, David S},
  journal={International Journal of Molecular Sciences},
  volume={22},
  number={16},
  pages={8931},
  year={2021},
  publisher={MDPI}
}

@article{jingjing2023three,
  title={Three-input logic gate based on DNA strand displacement reaction},
  author={Jingjing, MA},
  journal={Scientific Reports},
  volume={13},
  number={1},
  pages={15210},
  year={2023},
  publisher={Nature Publishing Group UK London}
}

@article{zhou2016multifunctional,
  title={Multifunctional graphene/DNA-based platform for the construction of enzyme-free ternary logic gates},
  author={Zhou, Chunyang and Liu, Dali and Wu, Changtong and Dong, Shaojun and Wang, Erkang},
  journal={ACS Applied Materials \& Interfaces},
  volume={8},
  number={44},
  pages={30287--30293},
  year={2016},
  publisher={ACS Publications}
}

@article{zhang2023implementing,
  title={Implementing logic gates by DNA crystal engineering},
  author={Zhang, Cuizheng and Paluzzi, Victoria E and Sha, Ruojie and Jonoska, Natasha and Mao, Chengde},
  journal={Advanced Materials},
  volume={35},
  number={33},
  pages={2302345},
  year={2023},
  publisher={Wiley Online Library}
}

@article{allemani2018global,
  title={Global surveillance of trends in cancer survival 2000--14 (CONCORD-3): analysis of individual records for 37 513 025 patients diagnosed with one of 18 cancers from 322 population-based registries in 71 countries},
  author={Allemani, Claudia and Matsuda, Tomohiro and Di Carlo, Veronica and Harewood, Rhea and Matz, Melissa and Nik{\v{s}}i{\'c}, Maja and Bonaventure, Audrey and Valkov, Mikhail and Johnson, Christopher J and Est{\`e}ve, Jacques and others},
  journal={The Lancet},
  volume={391},
  number={10125},
  pages={1023--1075},
  year={2018},
  publisher={Elsevier}
}

@article{zandvakili2019cell,
  title={Cell-free DNA testing: future applications in gastroenterology and hepatology},
  author={Zandvakili, Inuk and Lazaridis, Konstantinos N},
  journal={Therapeutic Advances in Gastroenterology},
  volume={12},
  pages={1756284819841896},
  year={2019},
  publisher={SAGE Publications Sage UK: London, England}
}

@article{shieh2020advances,
  title={Advances in the Genetic Testing of Neuromuscular Diseases.},
  author={Shieh, Perry B},
  journal={Neurologic Clinics},
  volume={38},
  number={3},
  pages={519--528},
  year={2020}
}

@article{liu2017regulating,
  title={Regulating DNA self-assembly by DNA--surface interactions},
  author={Liu, Longfei and Li, Yulin and Wang, Yong and Zheng, Jianwei and Mao, Chengde},
  journal={ChemBioChem},
  volume={18},
  number={24},
  pages={2404--2407},
  year={2017},
  publisher={Wiley Online Library}
}

@article{oishi2018comparative,
  title={Comparative study of DNA circuit system-based proportional and exponential amplification strategies for enzyme-free and rapid detection of miRNA at room temperature},
  author={Oishi, Motoi},
  journal={ACS omega},
  volume={3},
  number={3},
  pages={3321--3329},
  year={2018},
  publisher={ACS Publications}
}

@article{lv2021biocomputing,
  title={Biocomputing based on DNA strand displacement reactions},
  author={Lv, Hui and Li, Qian and Shi, Jiye and Fan, Chunhai and Wang, Fei},
  journal={ChemPhysChem},
  volume={22},
  number={12},
  pages={1151--1166},
  year={2021},
  publisher={Wiley Online Library}
}

@article{wang2023parallel,
  title={Parallel molecular computation on digital data stored in DNA},
  author={Wang, Boya and Wang, Siyuan Stella and Chalk, Cameron and Ellington, Andrew D and Soloveichik, David},
  journal={Proceedings of the National Academy of Sciences},
  volume={120},
  number={37},
  pages={e2217330120},
  year={2023},
  publisher={National Academy of Sciences}
}

@article{chen2023dna,
  title={DNA strand displacement based computational systems and their applications},
  author={Chen, Congzhou and Wen, Jinda and Wen, Zhibin and Song, Sijie and Shi, Xiaolong},
  journal={Frontiers in Genetics},
  volume={14},
  pages={1120791},
  year={2023},
  publisher={Frontiers Media SA}
}

@article{xu2022graph,
  title={Graph computation using algorithmic self-assembly of DNA molecules},
  author={Xu, Jin and Chen, Congzhou and Shi, Xiaolong},
  journal={ACS Synthetic Biology},
  volume={11},
  number={7},
  pages={2456--2463},
  year={2022},
  publisher={ACS Publications}
}

@article{li2015implementation,
  title={Implementation of arithmetic functions on a simple and universal molecular beacon platform},
  author={Li, Hailong and Guo, Shaojun and Liu, Qinghui and Qin, Lidong and Dong, Shaojun and Liu, Yaqing and Wang, Erkang},
  journal={Advanced Science},
  volume={2},
  number={5},
  pages={1500054},
  year={2015},
  publisher={Wiley Online Library}
}

@article{han20188,
  title={8-bit adder and subtractor with domain label based on DNA strand displacement},
  author={Han, Weixuan and Zhou, Changjun},
  journal={Molecules},
  volume={23},
  number={11},
  pages={2989},
  year={2018},
  publisher={MDPI}
}

@article{li2016dna,
  title={DNA based arithmetic function: a half adder based on DNA strand displacement},
  author={Li, Wei and Zhang, Fei and Yan, Hao and Liu, Yan},
  journal={Nanoscale},
  volume={8},
  number={6},
  pages={3775--3784},
  year={2016},
  publisher={Royal Society of Chemistry}
}

@article{su2019high,
  title={High-efficiency and integrable DNA arithmetic and logic system based on strand displacement synthesis},
  author={Su, Haomiao and Xu, Jinglei and Wang, Qi and Wang, Fuan and Zhou, Xiang},
  journal={Nature communications},
  volume={10},
  number={1},
  pages={5390},
  year={2019},
  publisher={Nature Publishing Group UK London}
}

@article{wang2020implementing,
  title={Implementing digital computing with DNA-based switching circuits},
  author={Wang, Fei and Lv, Hui and Li, Qian and Li, Jiang and Zhang, Xueli and Shi, Jiye and Wang, Lihua and Fan, Chunhai},
  journal={Nature communications},
  volume={11},
  number={1},
  pages={121},
  year={2020},
  publisher={Nature Publishing Group UK London}
}

@article{xie2022scaling,
  title={Scaling up multi-bit DNA full adder circuits with minimal strand displacement reactions},
  author={Xie, Nuli and Li, Mingqiang and Wang, Yue and Lv, Hui and Shi, Jiye and Li, Jiang and Li, Qian and Wang, Fei and Fan, Chunhai},
  journal={Journal of the American Chemical Society},
  volume={144},
  number={21},
  pages={9479--9488},
  year={2022},
  publisher={ACS Publications}
}

@article{xu2025dna,
  title={Dna coding theory and algorithms},
  author={Xu, Jin and Liu, Wenbin and Zhang, Kai and Zhu, Enqiang},
  journal={Artificial Intelligence Review},
  volume={58},
  number={6},
  pages={178},
  year={2025},
  publisher={Springer}
}

@article{ozawa2017role,
  title={The role of CH/$\pi$ interactions in the high affinity binding of streptavidin and biotin},
  author={Ozawa, Motoyasu and Ozawa, Tomonaga and Nishio, Motohiro and Ueda, Kazuyoshi},
  journal={Journal of Molecular Graphics and Modelling},
  volume={75},
  pages={117--124},
  year={2017},
  publisher={Elsevier}
}

@article{zhang2015reconfigurable,
  title={Reconfigurable and resettable arithmetic logic units based on magnetic beads and DNA},
  author={Zhang, Siqi and Wang, Kun and Huang, Congcong and Sun, Ting},
  journal={Nanoscale},
  volume={7},
  number={48},
  pages={20749--20756},
  year={2015},
  publisher={Royal Society of Chemistry}
}

@article{park2012simple,
  title={Simple and universal platform for logic gate operations based on molecular beacon probes},
  author={Park, Ki Soo and Seo, Myung Wan and Jung, Cheulhee and Lee, Joon Young and Park, Hyun Gyu},
  journal={Small},
  volume={8},
  number={14},
  pages={2203--2212},
  year={2012},
  publisher={Wiley Online Library}
}

@article{goni2019high,
  title={High-performance biocomputing in synthetic biology--integrated transcriptional and metabolic circuits},
  author={Go{\~n}i-Moreno, Angel and Nikel, Pablo I},
  journal={Frontiers in bioengineering and biotechnology},
  volume={7},
  pages={40},
  year={2019},
  publisher={Frontiers Media SA}
}

@article{yang2021dna,
  title={DNA logic circuits for multiple tumor cells identification using intracellular MicroRNA molecular bispecific recognition},
  author={Yang, Qiqi and Yang, Fan and Dai, Wenhao and Meng, Xiangdan and Wei, Wei and Cheng, Yaru and Dong, Jinhong and Lu, Huiting and Dong, Haifeng},
  journal={Advanced healthcare materials},
  volume={10},
  number={21},
  pages={2101130},
  year={2021},
  publisher={Wiley Online Library}
}

@article{qian2011scaling,
  title={Scaling up digital circuit computation with DNA strand displacement cascades},
  author={Qian, Lulu and Winfree, Erik},
  journal={science},
  volume={332},
  number={6034},
  pages={1196--1201},
  year={2011},
  publisher={American Association for the Advancement of Science}
}

@article{qian2011neural,
  title={Neural network computation with DNA strand displacement cascades},
  author={Qian, Lulu and Winfree, Erik and Bruck, Jehoshua},
  journal={nature},
  volume={475},
  number={7356},
  pages={368--372},
  year={2011},
  publisher={Nature Publishing Group UK London}
}

@article{evans2017physical,
  title={Physical principles for DNA tile self-assembly},
  author={Evans, Constantine G and Winfree, Erik},
  journal={Chemical Society Reviews},
  volume={46},
  number={12},
  pages={3808--3829},
  year={2017},
  publisher={Royal Society of Chemistry}
}

@article{woods2019diverse,
  title={Diverse and robust molecular algorithms using reprogrammable DNA self-assembly},
  author={Woods, Damien and Doty, David and Myhrvold, Cameron and Hui, Joy and Zhou, Felix and Yin, Peng and Winfree, Erik},
  journal={Nature},
  volume={567},
  number={7748},
  pages={366--372},
  year={2019},
  publisher={Nature Publishing Group UK London}
}

@article{sarraf2023modular,
  title={Modular reconfiguration of DNA origami assemblies using tile displacement},
  author={Sarraf, Namita and Rodriguez, Kellen R and Qian, Lulu},
  journal={Science Robotics},
  volume={8},
  number={77},
  pages={eadf1511},
  year={2023},
  publisher={American Association for the Advancement of Science}
}

@article{amir2014universal,
  title={Universal computing by DNA origami robots in a living animal},
  author={Amir, Yaniv and Ben-Ishay, Eldad and Levner, Daniel and Ittah, Shmulik and Abu-Horowitz, Almogit and Bachelet, Ido},
  journal={Nature nanotechnology},
  volume={9},
  number={5},
  pages={353--357},
  year={2014},
  publisher={Nature Publishing Group UK London}
}

@article{chatterjee2017spatially,
  title={A spatially localized architecture for fast and modular DNA computing},
  author={Chatterjee, Gourab and Dalchau, Neil and Muscat, Richard A and Phillips, Andrew and Seelig, Georg},
  journal={Nature nanotechnology},
  volume={12},
  number={9},
  pages={920--927},
  year={2017},
  publisher={Nature Publishing Group UK London}
}

@article{bui2018localized,
  title={Localized DNA hybridization chain reactions on DNA origami},
  author={Bui, Hieu and Shah, Shalin and Mokhtar, Reem and Song, Tianqi and Garg, Sudhanshu and Reif, John},
  journal={ACS nano},
  volume={12},
  number={2},
  pages={1146--1155},
  year={2018},
  publisher={ACS Publications}
}

@article{li2025programming,
  title={Programming Directional Strand Polymerization on DNA Origami for Logic Computing},
  author={Li, Zimu and Hao, Yaya and Tang, Yuqing and Sun, Chenyun and Cheng, Jianing and Xia, Kai and Wang, Fei and Zuo, Xiaolei and Fan, Chunhai and Lv, Hui},
  journal={Small Structures},
  pages={2500220},
  year={2025},
  publisher={Wiley Online Library}
}

@article{song2018improving,
  title={Improving the performance of DNA strand displacement circuits by shadow cancellation},
  author={Song, Tianqi and Gopalkrishnan, Nikhil and Eshra, Abeer and Garg, Sudhanshu and Mokhtar, Reem and Bui, Hieu and Chandran, Harish and Reif, John},
  journal={ACS nano},
  volume={12},
  number={11},
  pages={11689--11697},
  year={2018},
  publisher={ACS Publications}
}

@article{sterin2025thermodynamically,
  title={A Thermodynamically Favoured Molecular Computer: Robust, Fast, Renewable, Scalable},
  author={St{\'e}rin, Tristan and Eshra, Abeer and Adio, Janet and Evans, Constantine Glen and Woods, Damien},
  journal={bioRxiv},
  pages={2025--07},
  year={2025},
  publisher={Cold Spring Harbor Laboratory}
}

\noindent\textbf{Acknowledgements}\\
This work was supported by the National Major Scientific Instrument and Equipment Development Project of the National Natural Science Foundation of China (No. 62427811) and the National Natural Science Foundation of China (No. 62272115). The authors declare no competing financial interest.\\

\noindent\textbf{Author contributions}\\
Conceptualization, E.Z. and P.Q.; Investigation, E.Z., X.L. and C.L.; Methodology, J.X., X.L. and P.Q.; Supervision, X.L. and C.L.; Writing-original draft, E.Z, P.Q. and  X.L.; Writing-review and editing, E.Z. and C.L. All authors have read and agreed to the published version of the manuscript.\\


%

\end{document}